\documentclass[twocolumn,prl,floatfix,showpacs,preprintnumbers,nofootinbib]{revtex4}
\usepackage[utf8x]{inputenc}
\usepackage{mathrsfs}
\usepackage{amssymb}
\usepackage{amsmath}
\usepackage{graphicx}

\usepackage[caption = false]{subfig}

\usepackage{xcolor}
\definecolor{lcolor}{rgb}{0.5,0,0}
\definecolor{citcolor}{rgb}{0,0.3,0.0}

\usepackage[breaklinks,colorlinks,urlcolor=blue,citecolor=citcolor,linkcolor=lcolor]{hyperref}
\usepackage{mciteplus}

\newcommand{\rt}{{\mathbf{r}}}
\newcommand{\xt}{{\mathbf{x}}}
\newcommand{\bt}{{\mathbf{b}}}
\newcommand{\bti}{{\mathbf{b}_{i}}}
\newcommand{\yt}{{\mathbf{y}}}

\newcommand{\Deltat}{{\boldsymbol{\Delta}}}

\newcommand{\ud}{\, \mathrm{d}}

\newcommand{\tr}{\, \mathrm{Tr} \, }

\newcommand{\nc}{{N_\mathrm{c}}}

\newcommand{\gev}{\ \textrm{GeV}}
\newcommand{\fm}{\ \textrm{fm}}

\newcommand{\as}{\alpha_{\mathrm{s}}}

\newcommand{\sigmap}{{ \sigma^\textrm{p}_\textrm{dip} }}

\newcommand{\dsigmap}{{\frac{\ud \sigma^\textrm{p}_\textrm{dip}}{\ud^2 \bt}}}

\newcommand{\xpom}{{x_\mathbb{P}}}

\newcommand{\der}{\mathrm{d}}

\begin{document}

\author{Heikki M\"antysaari}
\affiliation{
Physics Department, Brookhaven National Laboratory, Upton, NY 11973, USA
}

\author{Bj\"orn Schenke}
\affiliation{
Physics Department, Brookhaven National Laboratory, Upton, NY 11973, USA
}

\title{
Evidence of strong proton shape fluctuations from incoherent diffraction
}

\pacs{24.85.+p, 13.60.-r}

\preprint{}

\begin{abstract}
We show within the saturation framework that measurements of exclusive vector meson production at high energy provide evidence for strong geometric fluctuations of the proton. In comparison, the effect of saturation scale and color charge fluctuations is weak. This knowledge will allow detailed future measurements of the incoherent cross section to tightly constrain the fluctuating geometry of the proton as a function of the parton momentum fraction $x$.
\end{abstract}

\maketitle

\paragraph{Introduction}

It is of fundamental interest to understand the geometric structure of the proton. This includes both its average size, shape and fluctuations.
The conventional way to access proton structure is via deep inelastic scattering (DIS) measurements where a (virtual) photon scatters off the proton. For example, the average proton density described in terms of parton distribution functions has been extracted accurately by H1 and ZEUS experiments at HERA~\cite{Aaron:2009aa,Abramowicz:2015mha}. One of the striking findings of these measurements is the rapid growth of the gluon density at small 
Bjorken $x$ or, equivalently, at high energy. 

At very high energy the Color Glass Condensate (CGC) effective theory of quantum chromodynamics (QCD) provides a natural framework to
describe the scattering processes \cite{Iancu:2003xm,Gelis:2010nm}. In fact, the accurate description of the proton structure function data has been a 
crucial test for the CGC picture \cite{Albacete:2010sy,Rezaeian:2012ji,Lappi:2013zma}. 

Accessing fluctuations of the geometric shape of the proton\footnote{In this work geometric shape refers to the spatial gluon distribution of the proton.} is a difficult task.
Two possible venues could allow progress in this direction:
On the one hand, high multiplicity proton-proton and proton-nucleus scattering experiments 
performed at the Relativistic Heavy Ion collider (RHIC) and Large Hadron Collider (LHC) have 
revealed striking collective phenomena \cite{Khachatryan:2010gv,Abelev:2012ola,Aad:2012gla,Adare:2013piz} (see \cite{Dusling:2015gta} for a recent review).
The interpretation that these effects are due to the collective expansion of the system, which is sensitive to the initial 
fluctuating geometry as in heavy-ion collisions, opens up the possibility to access the fluctuating proton shape in these experiments \cite{Coleman-Smith:2013rla,Schenke:2014zha,Aad:2015zza}.

Alternatively, diffractive DIS~\cite{PhysRev.120.1857,Miettinen:1978jb}, especially exclusive vector meson production, can provide insight into the average proton density distribution and its fluctuations. 

In such events, a vector meson (e.g. $J/\Psi$) is produced with no exchange of net color charge with the proton.
If the target proton remains intact (\emph{coherent diffraction}) the average color charge density profile is probed.  If, on the other hand, the proton breaks up, one becomes sensitive to event-by-event fluctuations~\cite{Miettinen:1978jb,Frankfurt:1993qi,Frankfurt:2008vi,Dominguez:2008aa,Caldwell:2009ke,Marquet:2009vs}. Diffractive processes with protons and heavier targets have been studied in perturbative QCD~\cite{Barone:2002cv} 
and the CGC framework \cite{Kovchegov:1999ji,kuroda:2005by,Goncalves:2005yr,Kowalski:2008sa,Caldwell:2009ke,Lappi:2010dd,Lappi:2013am,Lappi:2014foa}. 
In this letter, we calculate diffractive vector meson production at small $x$ within an event-by-event CGC framework. We constrain the amount of proton shape fluctuations using comparisons to available incoherent HERA data.
 A future electron ion collider \cite{Boer:2011fh,Accardi:2012qut} has the potential to provide much more precise data in a wider kinematical range~\cite{Toll:2012mb,Accardi:2012qut}, which can provide even stronger constraints.

\paragraph{Diffractive DIS in the dipole picture}
We study the exclusive production of a vector meson $V$ with momentum $P_V$ in deep inelastic scattering:
\begin{equation}
	l(\ell) + p(P) \rightarrow l'(\ell') + p'(P') + V(P_V),
\end{equation}
where $\ell$ and $\ell'$ are the lepton ($l$) momenta in the initial and final state, respectively, while $P$ and $P'$ are the initial and final proton ($p$) momenta. The Lorentz invariant quantities that describe the kinematics are
$Q^2=-q^2=-(\ell-\ell')^2$, $t \equiv -(P'-P)^2$, $\xpom \equiv (P-P')\cdot q/(P \cdot q) = (M^2+Q^2-t)/(W^2+Q^2-m_N^2)$.
Here $m_N$ is the proton mass, $M$ is the mass of the produced vector meson and $W^2=(P+q)^2$ is the total center-of-mass energy squared of the virtual photon-proton scattering. In diffractive DIS $\xpom$ plays the role of Bjorken-$x$ in DIS, and it corresponds to the longitudinal momentum fraction transferred from the target proton.

In the dipole picture diffractive vector meson production involves the fluctuation of a virtual photon, emitted by the lepton, into a quark-antiquark color dipole. This dipole then scatters elastically off the target hadron without exchanging net color charge, and finally forms the diffractively produced vector meson, in our analysis a $J/\Psi$ meson.

Following Ref.~\cite{Kowalski:2006hc} the scattering amplitude for the diffractive vector meson production can be written as
\begin{multline}
\label{eq:diff_amp}
 A^{\gamma^* p \to J/\Psi p}_{T,L}(\xpom,Q^2, \boldsymbol{\Delta}) = i\int \der^2 \rt \int \der^2 \bt \int \frac{\der z}{4\pi}  \\ 
 \times (\Psi^*\Psi_V)_{T,L}(Q^2, \rt,z) \\
 \times e^{-i[\bt - (1-z)\rt]\cdot \boldsymbol{\Delta}}  \dsigmap(\bt,\rt,\xpom).
\end{multline}
All bold face characters are transverse two-vectors: $\rt$ is the vector between the dipole quark and anti-quark, $\bt$ the impact parameter (distance from the center of the proton to the center-of-mass of the dipole), $\Deltat=(P'-P)_\perp$ is the transverse momentum transfer. $z$ is the fraction of the virtual photon longitudinal momentum carried by the quark.
The indices $T,L$ refer to the photon polarization. In this work we only consider transverse polarization as we study photoproduction ($Q^2=0$) events. 
The large $J/\Psi$ mass provides the scale that suppresses non-perturbative contributions from large dipole sizes. The overlap between the virtual photon and vector meson wave functions is given by $\Psi^*\Psi_V$. The virtual photon wave function  $\Psi$, describing $\gamma^*\to q\bar q$ splitting, can be computed from perturbative QED, but the formation of a vector meson from a color dipole is a non-perturbative process and must be modeled. In this work, we use the Boosted Gaussian wave function parametrization \cite{Kowalski:2006hc}.

In coherent diffraction, the target proton remains intact and the cross section can be written as~\cite{Kowalski:2006hc}
\begin{equation}
\frac{\der \sigma^{\gamma^* p \to J/\Psi p}}{\der t} = \frac{1}{16\pi} \left| \langle A(\xpom,Q^2,\boldsymbol{\Delta}) \rangle \right|^2.
\end{equation}
The brackets $\langle \rangle$ refer to  an average over target configurations. As we are only interested in high energy scattering, the small-$x$ structure of the proton dominated by gluons is probed. The coherent cross section is sensitive to the average gluon density, as it is obtained from the averaged scattering amplitude. On the other hand, when the proton breaks up (but the event is still diffractive, and there is no exchange of color charge between the proton and the vector meson), the incoherent cross section is obtained as a variance  (see e.g. Refs.~\cite{Caldwell:2009ke,Lappi:2010dd}):
\begin{align}\label{eq:incoherent}
\frac{\der \sigma^{\gamma^* N \to J/\Psi N^*}}{\der t} = \frac{1}{16\pi} &\left( \left\langle \left|  A(\xpom,Q^2,\boldsymbol{\Delta})  \right|^2 \right\rangle \right. \notag\\ 
& ~~~ - \left. \left| \langle A(\xpom,Q^2,\boldsymbol{\Delta}) \rangle \right|^2 \right)\,.
\end{align} 

The dipole-target cross section $\sigmap$ encodes all the QCD dynamics of the scattering process. It is related to the imaginary part of the forward dipole-target scattering amplitude $N$ via the optical theorem:
\begin{equation}
\dsigmap(\bt,\rt,\xpom) = 2 N(\rt,\bt,\xpom).
\end{equation}
The dipole amplitude $N$ in principle satisfies the small-$x$ JIMWLK \cite{JalilianMarian:1996xn,JalilianMarian:1997jx,JalilianMarian:1997gr,Iancu:2001md} or Balitsky-Kovchegov (BK) \cite{Balitsky:1995ub,Kovchegov:1999yj} evolution equation. However, knowledge of the impact parameter dependence of $N$ is crucial in order to evaluate the diffractive scattering amplitude \eqref{eq:diff_amp}, and the impact parameter dependent JIMWLK and BK equations develop unphysical Coulomb tails that should be regulated by confinement scale physics~\cite{GolecBiernat:2003ym,Schlichting:2014ipa}. Thus, 
we choose to use the impact parameter dependent saturation model (IPsat), as well as the IP-Glasma model \cite{Schenke:2012wb,Schenke:2013dpa} to determine the dipole amplitude.

In the IPsat model the dipole cross section is given by~\cite{Kowalski:2003hm}
\begin{equation}\label{eq:unfactbt}
\dsigmap(\bt,\rt,\xpom)
 = 2\,\left[ 1 - \exp\left(- {\mathbf r}^2  F(\xpom,\rt) T_p(\bt)\right) 
\right].
\end{equation} Here $T_p(\bt)$ is the proton's spatial profile function
\begin{equation}\label{eq:Tp}
    T_p(\bt)=\frac{1}{2\pi B_p} e^{-\bt^2/(2B_p)}\,.
\end{equation}
We have checked that using an exponential distribution increases the coherent cross section for $|t|>1\,{\rm GeV}^2$, with only small changes for smaller $|t|$. For the effect of different profile functions see \cite{Gotsman:2001ne}.
The function $F$ is proportional to the 
DGLAP evolved gluon distribution~\cite{Bartels:2002cj},
\begin{equation}
F(\xpom, \rt^2) = 
\frac{ \pi^2 }{2 \nc} \as \left(\mu^2 \right) 
\xpom g\left(\xpom,\mu^2 \right),  
\label{eq:BEKW_F}
\end{equation}
with $\mu^2=\mu_0^2 + 4/\rt^2$. The proton width $B_p=4\gev^{-2}$, $\mu_0^2$ and the initial condition for the DGLAP evolution of the gluon distribution $\xpom g$ are parameters of the model. They were determined in a successful fit of the IPSat model to HERA DIS data in \cite{Rezaeian:2012ji}. We use a charm mass of $m_c=1.4 \gev$.

In the IP-Glasma model \cite{Schenke:2012wb}, the dipole amplitude at a given $\xpom$, $N(\xt-\yt,(\xt+\yt)/2,\xpom)=1-\tr(V(\xt) V^\dag(\yt))/N_c$ can be calculated directly from the Wilson lines $V(\xt)$ of the proton. They are obtained after sampling color charges $\rho(x^-,\xt)$ from the IPsat color charge distribution (proportional to the saturation scale $Q_s(\xpom)$, defined using the IPsat dipole amplitude \cite{Schenke:2012fw}) and solving the Yang-Mills equations for the gluon fields:
\begin{equation}\label{eq:wilson}
  V (\xt) = P \exp\left({-ig\int dx^{-} \frac{\rho(x^-,\xt)}{\boldsymbol{\nabla}^2+m^2} }\right)\,.
\end{equation}
Here $P$ indicates path ordering and $m$ is an infrared cutoff that will affect the proton size and consequently the diffractive cross sections. Calculations are performed on a lattice with spacing $a=0.02\,{\rm fm}$. We have checked that smaller lattice spacings do not alter the results. For more details see \cite{Schenke:2012fw}.

\paragraph{Phenomenological corrections}
In the IPsat model, the cross sections are obtained by replacing the dipole amplitude with its imaginary part. Correcting them by a factor of $(1+\beta^2)$, with $\beta = \tan \pi \lambda/2$ and $\lambda = (\der \ln A_{T,L}^{\gamma^* p \to J/\Psi p} )/(\der \ln 1/\xpom)$, accounts for the real part \cite{Kowalski:2006hc}. The real part correction is of the order of $10\%$ in the kinematical range considered in this work and depends weakly on $|t|$. 
In the IP-Glasma the dipole amplitude is used directly and we do not include this correction.

The skewedness correction takes into account the off-diagonal nature of the gluon distribution involved in the process. In the linearized approximation, where two gluons are exchanged with the target, these gluons carry different longitudinal momentum fractions $x_1$ and $x_2$ (to and from the target, to achieve a total momentum transfer of $\xpom$). In the high energy limit the dominant contribution is from the region $\xpom \approx x_1 \gg x_2$, and the off-diagonal gluon distribution function can be expressed by the diagonal gluon distribution $\xpom g(\xpom)$ corrected by a skewedness factor $R_g$~\cite{Shuvaev:1999ce, Martin:1999wb,Kowalski:2006hc} $R_g = 2^{2\lambda_g+3}\Gamma(\lambda_g+5/2)/(\sqrt{\pi}\,\Gamma(\lambda_g+4))$ with $\lambda_g = (\der \ln \xpom g(\xpom,\mu^2))/(\ln 1/\xpom)$.
The skewedness correction is numerically important, being of the order of $40\%$ in the kinematical region relevant to this work. For the IP-Glasma cross section the skewedness correction is approximated by that to the round IPsat proton. 

\paragraph{Fluctuating proton shape}
\label{sec:fluctuations}
If the proton does not fluctuate the variance of the amplitude \eqref{eq:diff_amp} vanishes and the incoherent cross section (\ref{eq:incoherent}) is zero. However, the incoherent 
diffractive cross section was measured by HERA to be significant~\cite{Breitweg:1999jy,Chekanov:2002rm,Aktas:2003zi,Alexa:2013xxa} (see also Ref.~\cite{TheALICE:2014dwa}).\footnote{H1 data in Ref.~\cite{Aktas:2003zi} is the total diffractive cross section, which at high-$|t|$ is purely incoherent to good accuracy.}

We include geometric fluctuations of the proton in the IPsat model by introducing a constituent quark model, in which the gluonic density of the proton is distributed around the three valence quarks. The physical picture here is that when moving towards small $x$, the large-$x$ valence quarks radiate gluons that will be localized around the original positions of the quarks.

The density profile of each constituent quark is assumed to be Gaussian
\begin{equation}
T_q(\bt) = \frac{1}{2\pi B_q} e^{-\bt^2/(2B_q)}\,,
\end{equation}
with width parameter $B_q$.
This corresponds to the replacement
\begin{equation}\label{eq:TqReplace}
T_p(\bt) \rightarrow \frac{1}{N_q} \sum_{i=1}^{N_q} T_q(\bt-\bti)
\end{equation}
in the dipole cross section Eq.\,\eqref{eq:unfactbt}, where $\bti$ are the transverse coordinates of the $N_q=3$ constituent quarks, sampled from a Gaussian distribution with width $B_{qc}$.

In the linear regime ($\rt^2 \ll 1/Q_s^2$) contributions to the proton structure function $F_2$ in the fluctuating case are exactly the same as in the original IPsat framework, which is constrained by HERA data. Modifications of large $\rt^2$ contributions are possible, but their effects are suppressed in diffractive $J/\Psi$ production.

In the IP-Glasma model we use the same modifications to include fluctuations. Explicitly, we use Eq.\,\eqref{eq:unfactbt} with the replacement \eqref{eq:TqReplace} to determine $Q_s(\bt,\xpom)$ and from that the spatial color charge density distribution.

\paragraph{Saturation scale fluctuations}
In addition to geometric fluctuations the proton saturation scale $Q_s$ can fluctuate event-by-event. These fluctuations can be expected to be localized in a transverse area $\sim 1/Q_s^2$. 

Experimentally observed multiplicity distributions and rapidity correlations in p+p collisions are best described when 
the saturation scale fluctuates according to \cite{McLerran:2015qxa,Bzdak:2015eii}
\begin{equation}
\label{eq:qsfluct}
P\left( \ln \left(Q_s^2 / \langle Q_s^2  \rangle \right) \right) = \frac{1}{\sqrt{2\pi}\sigma} \exp \left[- \frac{\ln^2 \left( Q_s^2/\langle Q_s^2\rangle  \right)}{2\sigma^2}\right],
\end{equation}
with $\sigma \sim 0.5$. 
In the constituent quark model a natural way to implement these fluctuations is to have the saturation scale of each of the quarks fluctuate independently. As $Q_s^2$ is proportional to $F(\xpom,\rt) T_p(\bt)$ in \eqref{eq:unfactbt}, these fluctuations can be implemented by modifying the normalization of the quark density function $T_q$. 

Alternatively, one can implement the fluctuations by dividing the transverse plane into a grid, and let $Q_s^2$ fluctuate in each grid cell. 
The natural scale for the size of the grid cells is the typical $1/Q_s^2$ (cf. \cite{Dumitru:2012yr}), which for the EIC and HERA kinematics we consider corresponds to  $a\sim 0.4 \fm$.

\paragraph{Results}
To study the effect of geometric fluctuations on the coherent and incoherent diffractive $J/\Psi$ production cross sections, we vary the proton size $B_{qc}$ (which determines the distribution of quark centers) and quark width parameter $B_q$ in the IPsat framework. 
\begin{figure}[tb]
\centering
		\includegraphics[width=0.5\textwidth]{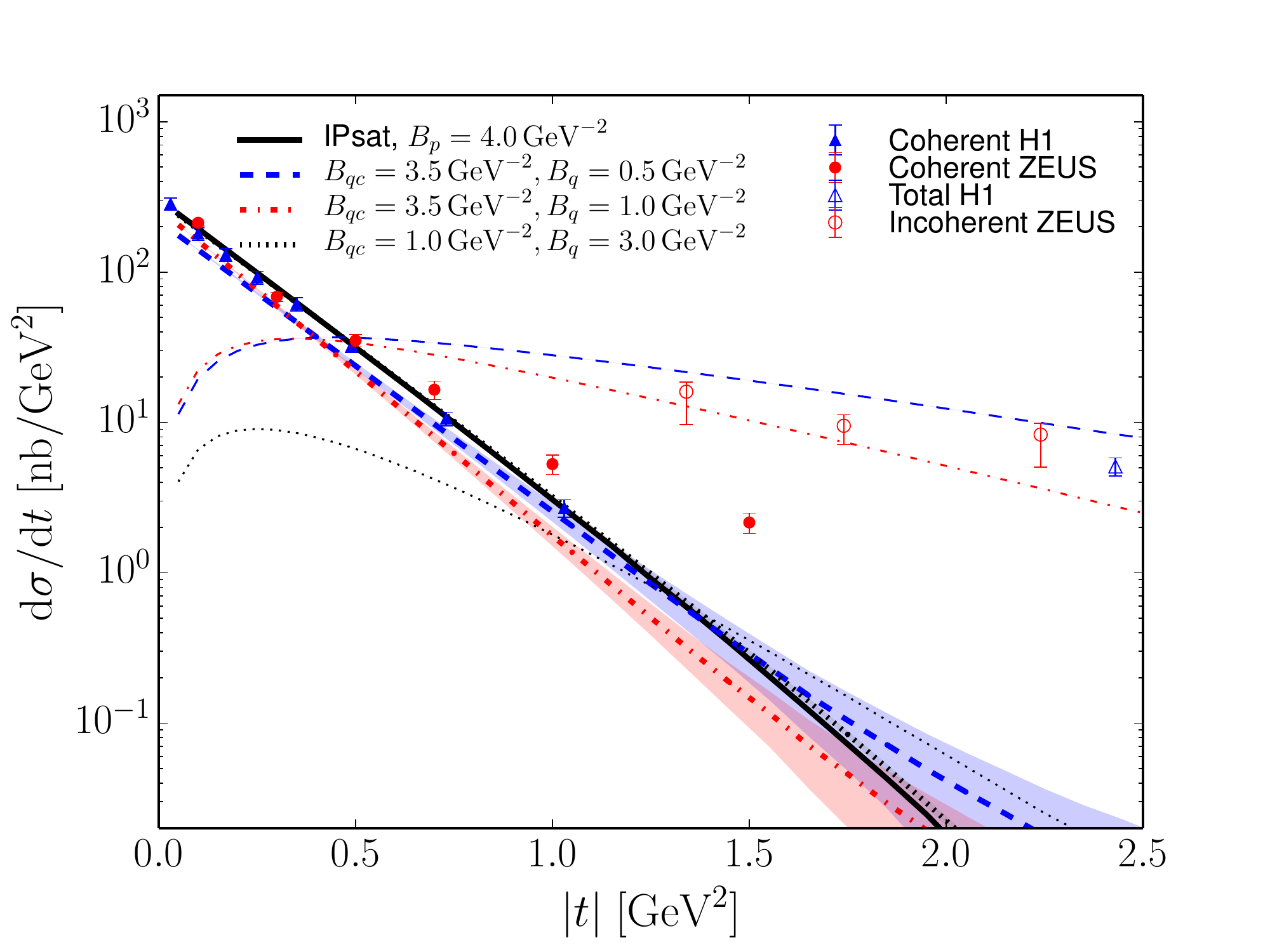} 
				\caption{Coherent (thick lines) and incoherent (thin lines) cross section as a function of $|t|$ compared with HERA data \cite{Aktas:2005xu,Chekanov:2002xi,Chekanov:2002rm,Aktas:2003zi}. The solid line is obtained without any fluctuations. The bands show statistical errors of the calculation.}
		\label{fig:bq-bp-10-35}
\end{figure}

Fig.~\ref{fig:bq-bp-10-35} shows both coherent and incoherent cross sections compared to HERA photoproduction data at $\langle W \rangle = 100\gev$ \cite{Aktas:2005xu,Chekanov:2002xi,Chekanov:2002rm,Aktas:2003zi}. With this kinematics one is sensitive to the proton structure at $x\approx 10^{-3}$. The incoherent cross section is extremely sensitive to the degree of geometric fluctuations of the proton.
While the smoother proton parametrized as $B_{qc}=1.0\gev^{-2}, B_q=3.0\gev^{-2}$ gives a good description of the coherent $J/\Psi$ production cross section, the incoherent cross section is largely underestimated (by more than an order of magnitude for $|t|\gtrsim 1\,\gev^2$). Increasing the amount of geometric fluctuations by using smaller quarks that are further apart on average ($B_{qc}=3.5\gev^{-2}, B_q=0.5\dots 1\gev^{-2}$), leads to an incoherent cross section compatible with the data, while maintaining a good description of the coherent $|t|$ spectrum. 
Consequently we also expect to maintain a good description of the $Q^2$ and $W$ dependence of the coherent $J/\Psi$ production cross section~\cite{Rezaeian:2012ji} and the agreement with the diffractive structure function data~\cite{Kowalski:2008sa} within the IPsat model.

Note that the average distance of a constituent quark from the center of the proton is $\sqrt{\langle r_q^2 \rangle} = \sqrt{ 2 B_{qc} }= 0.28 \fm$ for the smoother proton and $0.52\fm$ for the lumpy proton we consider.
We also show the conventional IPsat result, which has zero fluctuations and thus zero incoherent cross section.

\begin{figure}[tb]
\centering
\vspace{-2.2cm}
		\includegraphics[width=0.5\textwidth]{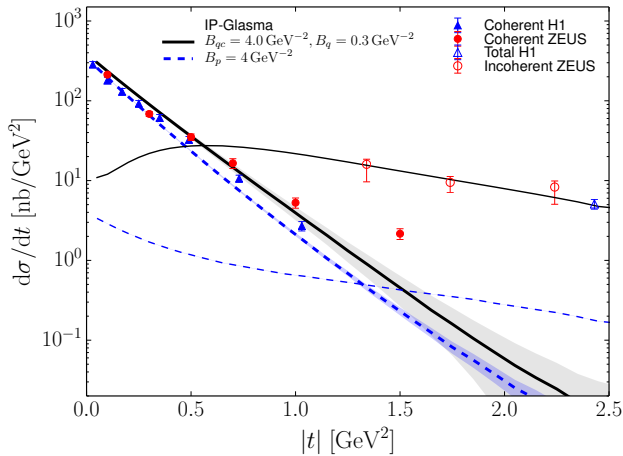} 
		\vspace{-2.31cm}
				\caption{Coherent and incoherent cross section as a function of $|t|$ calculated from the IP-Glasma framework compared with HERA data \cite{Aktas:2005xu,Chekanov:2002xi,Chekanov:2002rm,Aktas:2003zi}. The bands show statistical errors of the calculation.}
		\label{fig:spectra-glasma}
\end{figure}

In the IP-Glasma framework the additional color charge fluctuations produce a non-zero incoherent cross section even without geometric fluctuations. The effect of this kind of fluctuations on incoherent diffractive vector meson production was considered in \cite{Marquet:2010cf} in the Gaussian approximation and found to be suppressed as $1/N_c^2$.
The result for a round proton with $B_p=4 \gev^{-2}$ and $m=0.4\,\gev$ in Fig.~\ref{fig:spectra-glasma} shows that these fluctuations alone are not enough to describe the measured incoherent cross section. However, the IP-Glasma model combined with a constituent quark picture with parameters $B_{qc}=4 \gev^{-2}, B_q=0.3\gev^{-2}$, and $m=0.4\,\gev$ produces coherent and incoherent cross sections compatible with the data. This emphasizes the necessity of geometric fluctuations in a description of the transverse structure of the proton, which is in line with findings in p+A collisions~\cite{Schenke:2014zha}.

Note that even though the color charge density is sampled from a proton described by the IPsat model, in the IP-Glasma framework Coulomb tails are produced that are regulated by confinement scale physics implemented via the mass term $m$. These tails effectively increase the proton size, and when combined with the constituent quark model, weaken the fluctuations. It is the combination of $B_{qc}$, $B_q$ and $m$ that characterizes the degree of geometric fluctuations in the IP-Glasma framework. We have checked that reducing $m$ increases Coulomb tails and requires the reduction of $B_{qc}$ and $B_q$ to maintain agreement with the experimental data.

In the limit $t\to 0$ the incoherent cross section gets only a small contribution from geometric fluctuations. However, color charge fluctuations in the IP-Glasma model and possible $Q_s$ fluctuations are important in this limit. The geometric fluctuations start to dominate at $|t|\gtrsim 0.1\gev^2$. See Ref.~\cite{Miettinen:1978jb} for a more detailed discussion.

\begin{figure}[tb]
\centering
		\includegraphics[width=0.45\textwidth]{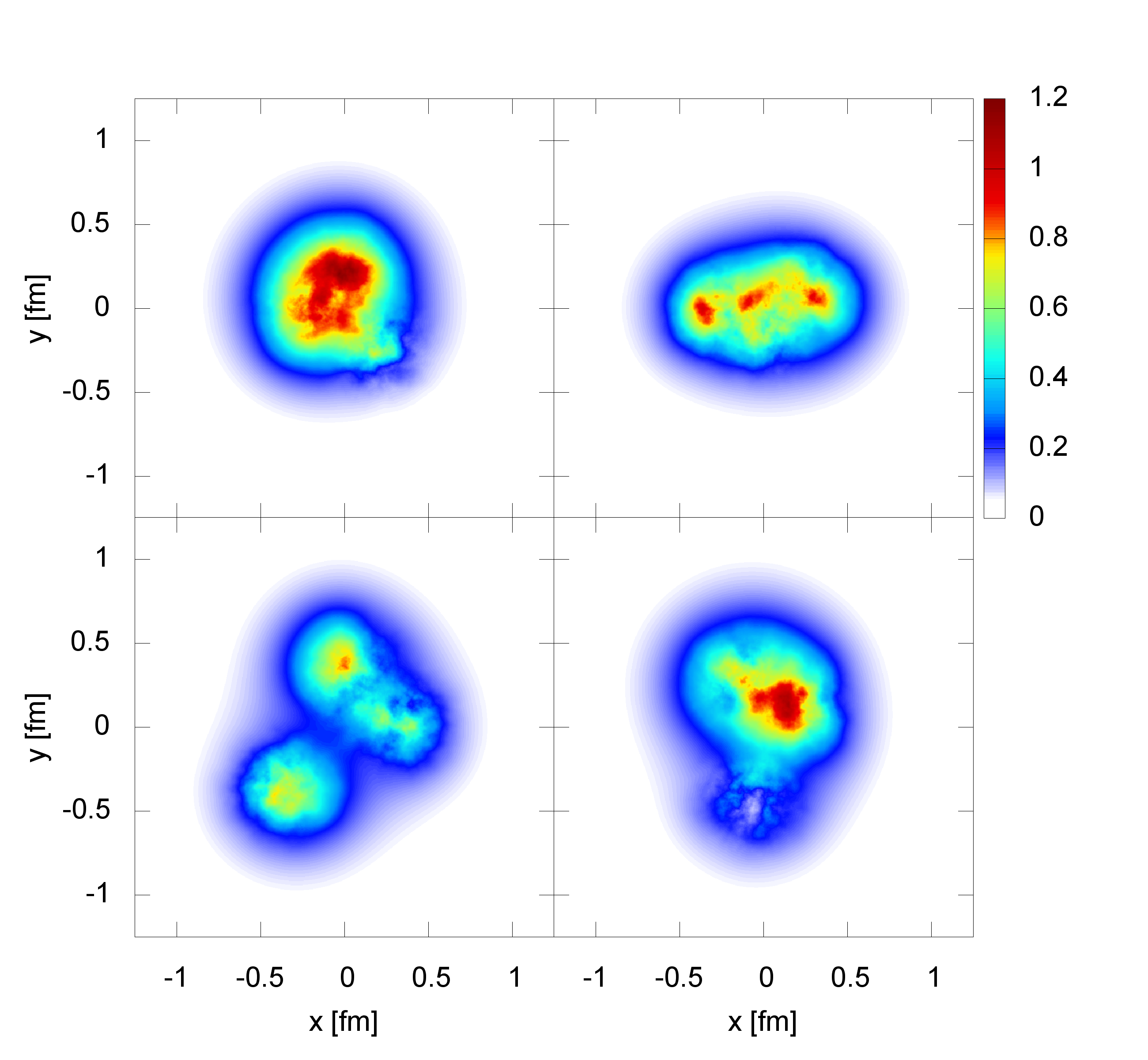} 
				\caption{Four configurations of the proton in the IP-Glasma model at $x\approx 10^{-3}$, represented by $1 - {\rm Re}(\tr V)/N_c$.}
		\label{fig:samples}
\vspace{0.8cm}
\end{figure}

Fig.\,\ref{fig:samples} shows example proton configurations in the IP-Glasma model with constituent quarks, demonstrating the strong shape variations required to achieve compatibility with experimental data. For simplicity, the quantity shown is $1 - {\rm Re}(\tr V)/N_c$.

Similar to the color charge fluctuations in the IP-Glasma framework, saturation scale fluctuations alone result in an incoherent cross section, which is orders of magnitude below the experimental data. The coherent and incoherent spectra obtained by combining geometric and $Q_s$ fluctuations on the level of
constituent quarks or on a grid with cell size $a=0.4\fm$ are almost indistinguishable. The incoherent cross section increases by $\sim 40\,\%$ for $|t|\gtrsim 0.3\gev^2$ when using $\sigma=0.5$. This increase is partially caused by the change of the expectation value of the log-normal probability distribution \eqref{eq:qsfluct} with the degree of fluctuations characterized by $\sigma$. Since this effect also increases the normalization of the coherent cross sections, it will be eliminated when we study the ratio of the cross sections.

\begin{figure}[tb]
\centering
\vspace{-2.3cm}
		\includegraphics[width=0.5\textwidth]{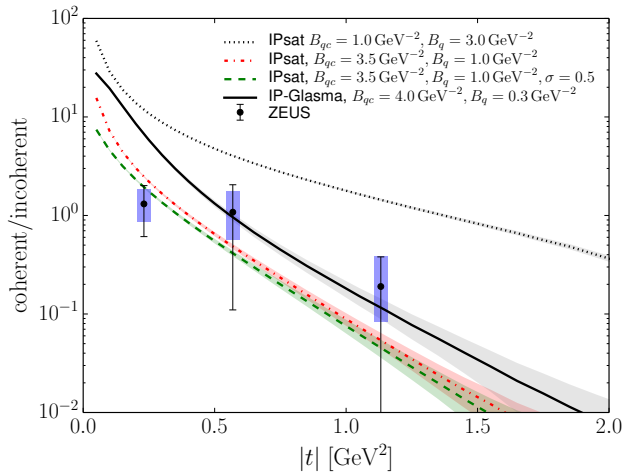} 
		\vspace{-2.6cm}
				\caption{Coherent cross section normalized by incoherent cross section compared with ZEUS data~\cite{Breitweg:1999jy}. The IPsat curve with $\sigma=0.5$ refers to the case where the saturation scales of the constituent quarks fluctuate following the distribution \eqref{eq:qsfluct}.}
		\label{fig:cohincoh}
\end{figure}

This ratio has been measured by the ZEUS collaboration starting from relatively small $|t|$~\cite{Breitweg:1999jy}. Fig.~\ref{fig:cohincoh} shows the ratio obtained with different parametrizations for the fluctuating structure of the proton. Again, a smoother proton ($B_{qc}=1.0\gev^{-2}, B_q=3.0\gev^{-2}$) is not in agreement with the experimental data, while a strongly fluctuating proton is compatible with the data. The effect of $Q_s$ fluctuations is small (within our statistical error) for $|t|\gtrsim 0.5\gev^2$.
Experimental errors are large around the intermediate $|t|\sim 1\gev^2$. This is exactly where a future electron ion collider can improve the situation by providing much more precise data to constrain the fluctuating proton shape.

\paragraph{Conclusions and Outlook}
In this letter we have shown that incoherent diffractive vector meson production provides a sensitive probe of the fluctuating shape of the proton. We implemented geometric fluctuations using a constituent quark model, and included realistic saturation scale and color charge fluctuations. We showed that the description of the incoherent cross section measured at HERA around $|t|\sim 0.5\dots 2 \gev^2$ requires strong geometric fluctuations. 
For a fixed proton geometry, saturation scale fluctuations constrained by LHC data \cite{McLerran:2015qxa} and color charge fluctuations as implemented in the IP-Glasma model \cite{Schenke:2012wb} underestimate the measured incoherent cross section by up to several orders of magnitude.
In the future, different models for the fluctuating proton geometry (see e.g. \cite{Bissey:2006bz}) can be implemented. However, we do not expect our results to be largely affected by the details of the model, as long as the degree of lumpiness remains the same.

More precise measurements from a future Electron Ion Collider will allow to tightly constrain the fluctuating shape of the proton at various values of Bjorken $x$. Our calculations can be improved by implementing QCD evolution via the JIMWLK equation~\cite{JalilianMarian:1996xn,JalilianMarian:1997jx,JalilianMarian:1997gr}, which can determine the $x$-dependence of the fluctuating proton shape~\cite{Schlichting:2014ipa}.

\paragraph*{Acknowledgments}
We thank E. Aschenauer, S. Schlichting, and T. Ullrich for discussions and T. Lappi and R. Venugopalan for valuable comments on the manuscript.
This work was supported under DOE Contract No. DE-SC0012704. This research used resources of the National Energy Research Scientific Computing Center, which is supported by the Office of Science of the U.S. Department of Energy under Contract No. DE-AC02-05CH11231. BPS acknowledges a DOE Office of Science Early Career Award.

\bibliographystyle{JHEP-2modM.bst}
\bibliography{./refs}

\end{document}